%% file: main.tex
\documentclass[onefignum,onetabnum]{siamonline190516}

\input{packages}

\headers{Sparse Tensor Decompositions via Chapel and C++/MPI Interoperability}{S. I. Geronimo Anderson and D. M. Dunlavy}

\title{Computing Sparse Tensor Decompositions via Chapel and C++/MPI Interoperability without Intermediate I/O}

\author{S. Isaac Geronimo Anderson\thanks{Sandia National Laboratories
		(\email{igeroni3@uoregon.edu/sgeroni@sandia.gov}, \email{dmdunla@sandia.gov})} \and Daniel M.
	Dunlavy\footnotemark[1] }

\ifpdf
\hypersetup{ pdftitle={Computing Sparse Tensor Decompositions via Chapel and C++/MPI Interoperability without Intermediate I/O}, pdfauthor={S. I. Anderson and D. M.
		Dunlavy} }
\fi

\begin{document}

\maketitle

\begin{abstract}
\input{abstract}
\end{abstract}



\capstartfalse
\begin{figure}[b!]
	\centering
	\includegraphics[width=\textwidth]{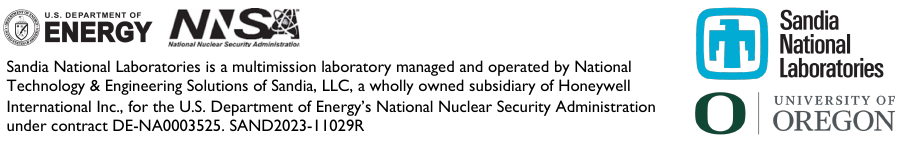}
\end{figure}
\capstarttrue

\input{content}

\section*{Acknowledgments}
\input{ack}

\bibliographystyle{siamplain}
\bibliography{refs}

\end{document}

%% file: packages.tex
\usepackage{csquotes}
\usepackage{bibentry}
\usepackage{cite}
\usepackage{epigraph}

\usepackage{bm}
\usepackage[mathscr]{euscript}
\usepackage{stmaryrd}
\SetSymbolFont{stmry}{bold}{U}{stmry}{m}{n} 
\usepackage[normalem]{ulem}
\usepackage{verbatim}
\usepackage{hyphenat} 
\usepackage{ulem}
\usepackage{cancel}

\usepackage{color}

\usepackage{listings}
\usepackage{hyperref}
\usepackage{cleveref}

\crefname{lstlisting}{tensor}{listings}
\Crefname{lstlisting}{Tensor}{Listings}

\usepackage[space=true]{accsupp}
\newcommand{\copyablespace}{\BeginAccSupp{method=hex,unicode,ActualText=00A0}\
	\EndAccSupp{}}

\usepackage{caption}
\usepackage{float}
\usepackage{graphicx}
\graphicspath{{./images/}}
\usepackage{subcaption}
\usepackage{tikz}
\usepackage{wrapfig}
\usepackage{pgfplots}
\pgfplotsset{compat=1.16}
\pgfplotsset{select coords between index/.style 2 args={ x filter/.code={
			\ifnum\coordindex<#1\fi
			\ifnum\coordindex>#2\fi } }}
\usepackage{booktabs}
\usepackage{ltablex}
\usepackage{multirow}
\usepackage{tabularx}
\usepackage{threeparttable}

\usepackage{adjustbox}

\usepackage{amsfonts,amsmath,amssymb,amsbsy}
\usepackage{mathtools}
\usepackage{etoolbox}

\usepackage{algorithm}
\usepackage[noend]{algpseudocode}
\Crefname{ALC@unique}{Line}{Lines} 
\algnewcommand{\LeftComment}[1]{\{#1\}\hfill}

\usepackage{enumitem}
\usepackage{siunitx}
\usepackage{url}

\usepackage{xparse}

\newcommand{\TODO}[2]{}
\newcommand{\EnableTODO}{%
	\renewcommand{\TODO}[2]{\textcolor{blue}{\textbf{\textit{[##1---##2]}}}}}
 \EnableTODO

%% file: abstract.tex
We extend an existing approach for efficient use of shared mapped memory across Chapel and C++ for graph data stored as 1-D arrays to sparse tensor data stored using a combination of 2-D and 1-D arrays. We describe the specific extensions that provide use of shared mapped memory tensor data for a particular C++ tensor decomposition tool called GentenMPI. We then demonstrate our approach on several real-world datasets, providing timing results that illustrate minimal overhead incurred using this approach. Finally, we extend our work to improve memory usage and provide convenient random access to sparse shared mapped memory tensor elements in Chapel, while still being capable of leveraging high performance implementations of tensor algorithms in C++.

%% file: content.tex
\section{Introduction}
\label{sec:intro}
We present several challenges and successes of sharing sparse tensor (i.e., multi-dimensional array) data, with no data duplication, between Chapel~\cite{Chamberlain11UserDefinedDomainMapsChapel} and a parallel program for computing sparse tensor decompositions that is written in C++ and uses the Message Passing Interface (MPI)~\cite{mpi} library to exchange data. Tensors are the higher-order generalization of matrices, and low-rank tensor decompositions (or factorizations) provide a useful tool for analyzing latent relationships in tensor data~\cite{KoldaBader09Review}. We focus here on computing low-rank canonical polyadic (CP) decompositions, for which several high-performance C++ implementations have been developed over the past decade~\cite{Devine20GentenMPI,Ranadive21AllAtOnce,Smith16Splatt,Teranishi20SparTen}.

While Chapel is portable and easy to use, we believe there is great benefit in leveraging existing high performance C++ libraries and applications from within a Chapel program.
If this were possible, then repeating years of effort in a Chapel reimplementation of a C++ library could be avoided, and Chapel could still be used for prototyping or for other operations on data.
One way to share data between Chapel programs and C++ applications is to write data to a file from Chapel, read the file into the C++ application, perform some computations, write the results to file from the C++ application, and then read the file into Chapel.
This incurs significant overhead which increases with the size of the data, in terms of file I/O for repeatedly writing and reading a file. It also incurs overhead in terms of memory bandwidth and storage for reading a copy of the file into the C++ application.

Previous work demonstrated sharing graphs represented as 1-D arrays between a Chapel program and a C++/MPI program, without writing the data to an intermediate file~\cite{McCrary22IntegratingChapel}.
The approach in that work effectively maintained data layout and locality, memory usage efficiency, and data-parallel computation capability. Here, we present two extensions of the previous 1-D work to multiple dimensions:

\begin{enumerate}
	\item \textbf{Coordinate array-based approach.} This approach directly implements a coordinate, or COO, format sparse tensor in a Chapel program using shared mapped memory. The sparse tensor is shared from Chapel to the C++/MPI program GentenMPI~\cite{Devine20GentenMPI} without duplicating the sparse data on disk or in memory. Crucially, the sparse data also adhere to GentenMPI's data layout and locality requirements. We demonstrate our use of this approach across several real-world data sets.
	\item \textbf{Sparse layout-based approach.} This approach implements a layout in Chapel for defining sparse multi-dimensional arrays. The coordinates are stored in shared mapped memory, and the coordinates can be added and used as array indices in the same way as a standard Chapel multi-dimensional array. This approach addresses several challenges presented by the coordinate array-based approach.
\end{enumerate}
%
To the best of our knowledge, our work is the first to extend the ability to share 1-D data between a Chapel program and a C++/MPI program to multiple dimensions.

\section{Background}
\label{sec:background}
Previous work enabled Chapel/C++ distributed-memory interoperability for graph algorithms whose graphs are represented using 1-D arrays (edge lists and optional weights).
McCrary \textit{et al.} demonstrated that it is possible to share data between Chapel programs and a C++/MPI graph analysis application, Grafiki (the successor of TriData~\cite{Wolf18TriData}), without duplicating the data on disk or in memory~\cite{McCrary22IntegratingChapel}.
They extended the Chapel \texttt{BlockDist} distribution to use shared mapped memory, via the low-level \texttt{shm\_open} and \texttt{mmap} Linux functions, to enable read/write access to distributed Chapel arrays by Grafiki MPI ranks.
Two one-dimensional (1-D) Chapel arrays store the pairs of vertices associated with each edge in the graph.
These arrays use the same Chapel domain constructed with the extended \texttt{BlockDist} distribution, hence automatically distributing the shared mapped memory arrays across the available Chapel locales (and, thus, corresponding MPI ranks).
Once the arrays are created and populated, the Chapel program writes a small metadata file including references to the shared mapped memory arrays.
The Chapel program then signals Grafiki using a flag passed via shared mapped memory that the arrays are ready for use in Grafiki.
The metadata file containing the shared mapped memory references is read from a single process on the MPI rank-0 node and is distributed to the other nodes. 
Each MPI rank opens its respective shared mapped memory array reference and wraps the corresponding memory in a \texttt{Kokkos::View}~\cite{Edwards13Kokkos}, which is the array abstraction format used by Grafiki.
Grafiki then performs the distributed graph analysis computations and signals Chapel via the shared mapped memory flag that the operation is finished.
The Chapel program can then perform further operations on the graph and/or on the results of the Grafiki computation.

\section{Methodology}
\label{sec:methodology}
The goal of this work is to extend the existing Chapel/C++ distributed-memory interoperability for sharing 1-D arrays between Chapel and C++/MPI programs to algorithms and programs that use sparse $d$-D tensors. We focus here on specific cases where $d \in \{3,4,5\}$, but our work covers the general case of $d \geq 1$.

GentenMPI stores sparse tensors in a coordinate, or COO, format that extends the standard Matrix Market format~\cite{MatrixMarket} used for sparse matrices. 
The GentenMPI COO format stores non-zero sparse tensor elements using a 2-D array for the coordinates (i.e., indices) and a 1-D array for the values. 
That is, for a $d$-dimensional sparse tensor with \texttt{nnz} non-zero elements, the coordinates are stored in a 2-D array of size $d \times \texttt{nnz}$, and the values are stored in a 1-D array of size \texttt{nnz}.
As a C++/MPI program, GentenMPI supports distributed operation but imposes constraints on the partitioning of the sparse tensor. Specifically, each array partition must contain only the elements belonging to a non-overlapping bounding box (sub-tensor) within the coordinate space of the sparse tensor.
For example, suppose a 3-D sparse tensor with dimensions $8 \times 6 \times 7$ using 1-based coordinates comprises two bounding boxes $A$ and $B$, where $A$ contains elements whose coordinates range from (1, 1, 1) to (4, 3, 3) and $B$ contains elements whose coordinates range from (5, 4, 4) to (8, 6, 7).
This scenario would require sparse tensor elements associated with $A$ to be stored together on one Chapel locale, and would require the same for elements associated with $B$.
GentenMPI requires the use of non-overlapping bounding boxes to minimize expensive inter-processor communication, thus maximizing overall computational performance.
We refer to this hereafter as the GentenMPI bounding box constraint.

\subsection{Coordinate Array-Based Approach}
\label{sec:methodology.coordinate}

Our first approach directly implements a 2-D coordinates array and a 1-D values array in Chapel, which together form a COO structure, while leveraging the previous Chapel/C++ interoperability work for graphs.
The \texttt{BlockDist} distribution allows 2-D and 1-D arrays, but the semantics of the 2-D coordinates array requires re-shaping the Chapel \texttt{Locales} array for proper distribution of the coordinates.
The semantics of the $d \times \texttt{nnz}$ 2-D coordinates array are such that the first dimension of the array---corresponding to the coordinates---should be considered indivisible, because all $d$ coordinates for a given non-zero element should be distributed together.
That is, if there is a non-zero element with coordinates $(i, j, k)$, it should not be the case that coordinates $i$ and $j$ belong to locale $A$ while coordinate $k$ belongs to locale $B$, when $B \neq A$.
The default behavior of \texttt{BlockDist} is to partition a 2-D array uniformly across both dimensions, and to distribute the partitions across the number of locales (\texttt{numLocales}).
Such behavior breaks the semantic indivisibility of the first dimension of the coordinates array.
But by reshaping the \texttt{Locales} array into a 2-D, $1 \times \texttt{numLocales}$ array, this leads to \texttt{BlockDist} partitioning the second dimension of the coordinates array---corresponding to the non-zero elements---as desired.

The GentenMPI bounding box constraint poses an issue for partitioning the coordinates array in Chapel, because the coordinates stored as elements in the array determine how the array should be partitioned, but Chapel partitions uniformly based on the dimensions of the 2-D coordinates array (i.e., $d \times \texttt{nnz}$), not the dimensions of the sparse tensor (e.g., $8 \times 6 \times 7$, as in the example above).
We address this constraint by performing a sparse tensor file analysis for determining partitioning information (a one-time cost), then setting the size of the coordinates array such that Chapel's uniform partitioning is compatible with the GentenMPI bounding box constraint.

\subsection{Sparse Layout-Based Approach}
\label{sec:methodology.sparse}

Our second approach implements a layout, or locale-specific domain map~\cite{Chamberlain11UserDefinedDomainMapsChapel}, in Chapel for defining multi-dimensional domains (hence multi-dimensional arrays) in Chapel.
The coordinates (or indices) of the domain are stored in shared mapped memory and hence can be shared from Chapel to a C++/MPI program.
Our layout extends Chapel's default \texttt{sparse subdomain} construct, which allows creating sparse arrays indexed by multi-dimensional coordinates.
Note that it is possible to use Chapel's default sparse subdomain with the extended ~\texttt{BlockDist}, but in that case only the values (not the coordinates) would reside in shared mapped memory.
It is not useful to be able to share sparse tensor values while not knowing the coordinates of those values.
Hence, in order to share a sparse tensor which uses a Chapel sparse subdomain, an extended layout that stores coordinates in shared mapped memory is required.
One example of an extended layout is the \texttt{LayoutCS} module in Chapel, which implements compressed sparse row storage for sparse domains where $d = 2$.
That is, \texttt{LayoutCS} is intended for use with sparse matrices and cannot be used directly in our work.



Chapel's default sparse subdomain stores coordinates in a COO-style format, meaning that the coordinates are stored in a 2-D array internal to the sparse subdomain data structure.
We cloned Chapel's sparse subdomain module and refactored this internal coordinates array to use shared mapped memory as in the \texttt{BlockDist} distribution described above.
Chapel sparse subdomains are initially empty, and indices must be added to the subdomain before array values can be stored at those indices.
Internally, Chapel uses a grow/shrink scheme for the 2-D coordinates array, and our approach uses \texttt{ftruncate}~\cite{ftruncate} and the GNU extension \texttt{mremap}~\cite{mremap} to adhere to the Chapel scheme while using shared mapped memory.

In considering the GentenMPI bounding box constraint, recall that Chapel partitions arrays uniformly based on the dimensions of the array. This means that for a sparse tensor which uses a Chapel sparse subdomain, the ``true'' dimensions of the array are the same as the sparse tensor's, and Chapel will partition the sparse tensor uniformly. In other words, our second approach leverages Chapel's (specifically \texttt{BlockDist}'s) partitioning to align with GentenMPI's bounding box constraint. This means that if our extended sparse layout can be used with the extended \texttt{BlockDist}, then Chapel's partitioning of the coordinates and values should adhere to GentenMPI's bounding box constraint.

One challenge that arose in the sparse layout-based approach that did not occur in the coordinate array-based approach was that of Chapel compile-time domain rank: Dimension, as in the number of modes of a sparse tensor (but not the size of each mode), must be known at compile time. This is because the dimension determines the rank of the associated Chapel domain, which must be known at compile time. We solved this challenge by wrapping our test program in a function with a dimension parameter (\texttt{param dimension}) and choosing a small number of dimensionalities (2-5) to support via a compound conditional. This solution appears to increase linearly the compile time and compiled executable size, but allows loading a variety of dimensions of sparse tensors (2-5) without recompiling.

\section{Experiments}
\label{sec:status}
We demonstrate the use of our coordinate array-based method (\textsection \ref{sec:methodology.coordinate}) for sparse tensor interoperability between Chapel and C++/MPI using several real-world sparse tensor datasets from the FROSTT~\cite{Smith17Frostt} tensor benchmark data repository. Table~\ref{tab:frostt_data} shows the details of the three tensor datasets used in the experiments presented here. Note that we do not demonstrate the use of our sparse layout-based method (\textsection \ref{sec:methodology.sparse}) because the current implementation is limited to use on a single locale only. As such, the experiments shown here are specifically those regarding our coordinate-based method.

\begin{table}[h!]
	\begin{minipage}{\columnwidth}
		{\small
			\begin{center}
				\begin{tabular}{lcc}
					\toprule
					Sparse Tensor & Tensor Dimension Sizes & \texttt{nnz} \\
					\midrule
					chicago-crime & $6.2K \times 24 \times 77 \times 32$ & $\sim 5.3$ M \\
					lbnl-network & $1.6K \times 4.2K \times 1.6K \times 4.2K \times 868K$ & $\sim 1.7$ M \\
					nell-2 & $12K \times 9K \times 29K$  & $\sim 77$ M \\
					\bottomrule
				\end{tabular}
			\end{center}
		}
	\end{minipage}
	\caption{Datasets from the FROSTT tensor benchmark repository used in experiments.}
	\label{tab:frostt_data}
\end{table}

We use GentenMPI's implementation of GCP-ADAM~\cite{Kolda20StochasticGradientsGCP} for computing the low-rank CP tensor decompositions, which is well suited for the FROSTT data, as it can compute decompositions on sparse count tensor data. We use GentenMPI's partitioning scheme, provided by the SPLATT tensor package~\cite{Smith16Splatt}, for computing the one-time cost of identifying non-overlapping bounding boxes as a function of the dataset and number of Chapel locales/MPI ranks. We vary the number of locales/ranks (4, 8, 16, 32, 64) used in the experiments to illustrate the performance characteristics over a range of computational resources. In each experiment, we run 5 iterations of the GCP-ADAM algorithm to compute an approximate rank-16 CP decomposition. We repeated each experiment five times, and recorded the mean and standard deviation for the runtime. All experiments were run using Chapel 1.24 on a Cray XC40 system, using up to 64 (of the 100 available) compute nodes, where each node has 32 cores and 128 GB of memory. Chapel 1.24 was used because it was the version of Chapel on our test system at the time of testing.



\begin{table*}[ht!]
	\begin{center}	
		{\footnotesize
			\begin{tabular}{lrcccc}
				\toprule
				Sparse tensor & Nodes & Chapel $\rightarrow$ C++ & Chapel $\rightarrow$ C++ & GCP-ADAM & GCP-ADAM \\
				&       & \texttt{shm\_open}/\texttt{mmap} & GentenMPI call & Chapel $\rightarrow$ C++ & C++ only \\
				&       & time (s) (Stdev)         & time (s) (Stdev) & time (s) (Stdev) & time (s) (Stdev) \\
				\midrule
				chicago-crime & 4 & 1.50e-03 (2e-05) & 5.76e-03 (2e-03) & 7.36e+00 (4e-02) & 7.20e+00 (3e-02) \\
				chicago-crime & 8 & 2.63e-03 (8e-04) & 5.06e-03 (5e-04) & 5.99e+00 (4e-02) & 5.86e+00 (5e-02) \\
				chicago-crime & 16 & 3.06e-03 (9e-05) & 1.74e-01 (2e-01) & 6.74e+00 (2e-02) & 6.62e+00 (3e-02) \\
				chicago-crime & 32 & 4.41e-03 (3e-05) & 6.70e-03 (6e-04) & 9.39e+00 (6e-02) & 9.41e+00 (3e-02) \\
				chicago-crime & 64 & 5.49e-03 (5e-05) & 1.04e-02 (4e-04) & 1.23e+01 (2e-02) & 1.22e+01 (5e-02) \\
				\midrule
				lbnl-network & 4 & 1.76e-03 (5e-04) & 5.43e-03 (9e-04) & 1.04e+03 (2e+01) & 1.07e+03 (9e+00) \\
				lbnl-network & 8 & 2.17e-03 (3e-04) & 4.48e-03 (4e-04) & 5.21e+02 (5e+00) & 5.26e+02 (8e-01) \\
				lbnl-network & 16 & 3.06e-03 (1e-04) & 1.02e-02 (8e-03) & 2.67e+02 (4e+00) & 2.69e+02 (6e-01) \\
				lbnl-network & 32 & 4.41e-03 (7e-05) & 6.36e-03 (3e-04) & 1.31e+02 (1e+00) & 1.31e+02 (6e-01) \\
				lbnl-network & 64 & 5.51e-03 (2e-05) & 1.31e-02 (5e-03) & 8.74e+01 (2e-01) & 8.71e+01 (5e-01) \\
				\midrule
				nell-2 & 4 & 1.49e-03 (7e-06) & 8.05e-03 (6e-04) & 5.06e+01 (2e-01) & 4.99e+01 (2e-01) \\
				nell-2 & 8 & 2.04e-03 (2e-05) & 1.30e-02 (8e-03) & 3.22e+01 (9e-02) & 3.20e+01 (1e-01) \\
				nell-2 & 16 & 2.91e-03 (4e-05) & 8.83e-03 (9e-04) & 2.03e+01 (1e-01) & 2.03e+01 (4e-02) \\
				nell-2 & 32 & 4.38e-03 (6e-05) & 9.76e-03 (2e-03) & 1.49e+01 (2e-01) & 1.49e+01 (3e-02) \\
				nell-2 & 64 & 5.49e-03 (3e-05) & 1.17e-02 (2e-03) & 1.19e+01 (2e-01) & 1.21e+01 (2e-01) \\
				\bottomrule
			\end{tabular}
		}
	\end{center}
	\caption{Timing results of Chapel and C++/MPI interoperability for computing low-rank sparse tensor decompositions over a range of number of Chapel locales/MPI ranks. Each column shows the mean (with standard deviations) for the runtime in seconds over five repeated experiments in the given configuration.}
	\label{tab:results}
\end{table*}

Note that we do not report timing results for file I/O at this time, as the methods in Chapel and GentenMPI are very different and thus not comparable. Instead, we focus on just the timing of the computations associated with shared mapped memory setup and computation across Chapel and C++/MPI versus C++/MPI alone.

Table~\ref{tab:results} presents timing results of these experiments. 
Column 3 (Chapel $\rightarrow$ C++ \texttt{shm\_open}/\texttt{mmap}) shows the time taken in allocating the shared mapped memory in Chapel and signaling GentenMPI to start. As shown, these times are quite small and appear to scale as the logarithm of the node count.  
Column 4 (Chapel $\rightarrow$ C++ GentenMPI call) shows the time taken to initialize the shared mapped memory in GentenMPI. These times are similarly small, except for an unusual outlier with \textit{chicago-crime} on 16 nodes---experiments are underway to determine why the GentenMPI call timing for this configuration is longer than for the others.
Finally, Columns 5 and 6 show the times for running GCP-ADAM using our Chapel $\rightarrow$ C++ interoperability and in C++ only (i.e., GentenMPI as a standalone program), respectively. These times are very similar to each other, suggesting non-significant amounts of overhead for our shared mapped memory approach. From a computational performance perspective, these are promising results.

\newpage
\section{Discussion}
\label{sec:discussion}
There are several research opportunities presented in the case of our coordinate array-based approach, including improved memory efficiency and general access to sparse tensor elements and operations from Chapel. For our sparse layout-based approach, distributed operation is the first main challenge to address, as our current implementation is limited to single-locale use.

Our coordinate array-based approach suffers from three challenges associated with memory efficiency:
\begin{enumerate}
	\item Chapel's \texttt{BlockDist} distribution partitions arrays uniformly, so the coordinates and values arrays are distributed in asymptotically equal portions across the locales.
	\item Sparse tensors generally exhibit irregular non-zero element patterns, which means that uniform bounding boxes for a sparse tensor generally will not contain equal portions of non-zero elements.
	\item The COO sparse tensor format (as used by sparse tensor decomposition libraries like GentenMPI) stores non-zero elements in dense arrays which will be uniformly partitioned by \texttt{BlockDist} based on their dense dimensions (i.e., $d \times \texttt{nnz}$).
\end{enumerate}

These three conditions lead to a situation where we must pad the dense arrays in order to guarantee that each locale has enough storage for its (generally irregularly large) number of bounding box non-zero elements.
This padding can lead to inefficient use of memory. For example, for the datasets used in our experiments here with poorly chosen bounding boxes, we could have some bounding boxes containing $O(1)$--$O(10)$ elements while others containing as much as $O(10^7)$ elements, leading to memory allocations of $O(10^7)$ across all locales. Attempts at addressing non-zero distribution irregularity through computational load balancing (e.g., in~\cite{Teranishi20SparTen}) has proven useful for such problems but alone could not address our inefficient use of memory via padding.
Hence, improving the memory efficiency in the coordinate array-based approach is a clear first research opportunity.

A second research opportunity regards the COO format itself, which does not allow for convenient random access to sparse tensor values.
This is because the coordinates of non-zero elements are stored in any order, and thus access would require a search over the entire coordinates array. Sorting the coordinates array---for example to support efficient computation (e.g., in~\cite{Phipps19SoftwareSparseTensor})---may help address the challenge of random access to sparse tensor values but at an additional cost of code complexity to support the many possible sorting schemes that can be applied to multidimensional arrays.

Our sparse layout-based approach addresses the shortcomings of the coordinate array-based approach in terms of memory efficiency and random access to sparse tensor values.
It improves memory efficiency by allowing the sparse domain (and associated sparse array) to grow as needed during initial sparse tensor construction.
That is, sparse indices can be added to a locale's local subdomain and the shared mapped memory region will grow to fit, without requiring a padded memory allocation.
A related benefit is that no sparse tensor file analysis is necessary for determining partition size information, as the number of local sparse indices does not need to be known a priori.
Our sparse layout-based approach also provides the ability to use Chapel's multi-dimensional array indexing for random access to sparse tensor values.
In terms of challenges, our sparse layout-based approach currently suffers from the lack of distributed operation.
The reason is due to time constraints rather than technical reasons.

While our sparse layout-based approach is improved with respect to both research opportunities presented by our coordinate array-based approach, the fact that it currently does not support multi-locale operation via \texttt{BlockDist} or connect Chapel with a C++/MPI application is a critical issue.
Hence, adding distributed operation to the sparse layout-based approach is a crucial first research opportunity.

In order to realize a solution from our sparse layout-based approach, we require the following:

\begin{enumerate}
	\item The ability to use our sparse layout as a \texttt{sparseLayoutType} when parameterizing \texttt{BlockDist}.
	\item \texttt{BlockDist} must allow multi-local usage with our sparse layout.
	\item Per-locale shared mapped memory file handles for local subdomains using our sparse layout, similarly to the previous work.
	\item Writing metadata files, similarly to the previous work.
	\item Connecting our Chapel program to a C++ program via metadata files and a shared mapped memory flag, similarly to the previous work.
\end{enumerate}

Note that several of these items are adaptations of the previous work, hence the majority of the design work is in the first two items.


\section{Conclusions}
\label{sec:conclusion}
In this report, we have presented our two approaches for extending existing Chapel/C++ distributed-memory interoperability to algorithms and programs that use sparse $d$-D tensors.
Our coordinate array-based approach showed promising results from a performance perspective. That is, the timing results for our approach were very similar to those of the traditional approach, without needing to duplicate data on disk or in memory.
Our sparse layout-based approach made progress towards addressing the shortcomings of the coordinate array-based approach, in terms of memory efficiency and random access to sparse tensor values.
We currently have a single-locale version of the sparse layout-based approach.
Finally, we have listed the remaining challenges for our sparse layout-based approach, and described our present thoughts on how to surmount those challenges.

%% file: ack.tex
We thank Karen Devine and Andrew Younge from Sandia National Laboratories for valuable help in extending their previous work on Chapel/C++ interoperability. We thank Jon Berry of Sandia National Laboratories for helpful discussions on the real-world problem motivations for Chapel/C++ interoperability. We thank Brad Chamberlain from Hewlett Packard Enterprise for help in understanding how to extend sparse capabilities within Chapel. We thank Jee Choi from the University of Oregon for suggestions regarding computational experiments and performance analyses.